\documentclass[conference]{IEEEtran}
\IEEEoverridecommandlockouts
\usepackage{cite}
\usepackage{amsmath,amssymb,amsfonts}
\usepackage{algorithmic}
\usepackage{graphicx}
\usepackage{textcomp}
\usepackage{xcolor}
\usepackage{subfigure}
\usepackage{algorithm, algorithmic}

\def\BibTeX{{\rm B\kern-.05em{\sc i\kern-.025em b}\kern-.08em
		T\kern-.1667em\lower.7ex\hbox{E}\kern-.125emX}}

\ifodd 0
\newcommand{\rev}[1]{{\color{blue}#1}} 
\newcommand{\com}[1]{\textbf{\color{green} (COMMENT: #1)}} 
\else
\newcommand{\rev}[1]{#1}

\newcommand{\com}[1]{}
\fi

\makeatletter
\newcommand\tinyv{\@setfontsize\tinyv{7pt}{9}}
\makeatother

\begin{document}

\title{Millimeter Wave Full-Duplex Networks: MAC Design and Throughput Optimization\\
}

\author{
\IEEEauthorblockN{Shengbo Liu\IEEEauthorrefmark{1}, Wen Wu\IEEEauthorrefmark{1}, Liqun Fu\IEEEauthorrefmark{2}, Kaige Qu\IEEEauthorrefmark{3},  Qiang Ye\IEEEauthorrefmark{4}, Weihua Zhuang\IEEEauthorrefmark{3}, and Sherman Shen\IEEEauthorrefmark{3}} 
\IEEEauthorblockA{\IEEEauthorrefmark{1}\textit{Frontier Research Center, Peng Cheng Laboratory, Shenzhen, China}} 
\IEEEauthorblockA{\IEEEauthorrefmark{2}\textit{School of Informatics, Xiamen University, Xiamen, China}} 
\IEEEauthorblockA{\IEEEauthorrefmark{3}\textit{Department of Electrical and Computer Engineering, University of Waterloo, Waterloo, Canada}} 
\IEEEauthorblockA{\IEEEauthorrefmark{4}\textit{Department of Computer Science, Memorial University of Newfoundland, St. John's, Canada}} 
Email: \{liushb, wuw02\}@pcl.ac.cn, liqun@xmu.edu.cn, \{k2qu, wzhuang, sshen\}@uwaterloo.ca, qiangy@mun.ca
}


\maketitle

\begin{abstract}

Full-duplex (FD) technique can remarkably boost the network capacity in the millimeter wave (mmWave) bands by enabling simultaneous transmission and reception. However, due to directional transmission and large bandwidth, the throughput and fairness performance of a mmWave FD network are affected by deafness and directional hidden-node (HN) problems and severe residual self-interference (RSI). To address these challenges, this paper proposes a directional FD medium access control protocol, named DFDMAC to support typical directional FD transmission modes by exploiting FD to transmit control frames to reduce signaling overhead. Furthermore, a novel busy-tone mechanism is designed to avoid deafness and directional HN problems and improve fairness of channel access. To reduce the impact of RSI on link throughput, we formulate a throughput maximization problem for different FD transmission modes and propose a power control algorithm to obtain the optimal transmit power. Simulation results show that the proposed DFDMAC can improve the network throughput and fairness by over $60\%$ and $32\%$, respectively, compared with the existing MAC protocol in IEEE 802.11ay. Moreover, the proposed power control algorithm can effectively enhance the network throughput.


\end{abstract}


\section{Introduction}
Benefiting from abundant spectrum resources, millimeter wave (mmWave) communication is capable of providing ultra-high data rate to facilitate a number of emerging applications, such as virtual reality and 8K video streaming  \cite{shen2021holistic}. To meet the fast-growing traffic demand in future networks, it is crucial to further enhance the spectrum efficiency of mmWave networks. Full-duplex (FD) communication
has the potential to double spectrum efficiency and capacity in the mmWave band by enabling simultaneous transmission and reception \cite{mmWaveFD}. Recent years have witnessed significant progress in mmWave FD system design \cite{mmFD_survey}, achieving substantial self-interference cancellation over large bandwidth. 
In 3GPP Release 17, FD has been adopted in integrated access and backhaul solution for deploying dense mmWave networks to reduce latency and improve spectrum efficiency \cite{10001634}. Furthermore, mmWave FD networks using unlicensed bands (e.g., 57-64 GHz) will play an essential role in the future 6G networks \cite{mmFD_survey}. Therefore, designing an FD-based medium access control (MAC) protocol that integrates well with existing mmWave networks using IEEE 802.11ay is critical for its standardization and deployment.




Existing FD MAC protocols are primarily developed for omnidirectional FD transmissions in sub-6 GHz bands \cite{dibaei2020full}. These protocols cannot be directly applied to mmWave FD networks that utilize directional transmissions through beamforming techniques to overcome high path loss \cite{wu2019fast}. Designing an MAC protocol in mmWave FD networks faces the following challenges.
 \begin{itemize}
	\item In contrast to a half-duplex (HD) network, a distributed millimeter-wave FD network requires multiple directional transmission modes to be supported. Using the existing RTS/CTS \rev{(request to send/ clear to send)} handshaking to establish FD transmission can incur significant overhead, which in turn leads to a degradation in throughput. \rev{It is challenging to coordinate FD transmission between nodes in a distributed manner while maintaining low overhead.} 
	\item Directional transmission with a narrow beam reduces signal coverage, which renders traditional carrier sensing mechanisms ineffective for accurately identifying channel state, resulting in deafness and directional hidden-node (HN) problems \cite{deaf_2021,Pielli}. 
Deafness problem occurs when a node cannot reply to a transmitter's request as it is beamformed towards another direction, and the transmitter treats this case as a collision and doubles its contention window, hence suffering unfair access \cite{deaf_2021}. HN problem arises when two nodes initiate transmissions to the same receiver simultaneously without sensing each other \cite{Pielli}. 
	\item Directional FD transmission link suffers from severe residual SI, which cannot be completely canceled due to large bandwidth, such as 2.16 GHz in IEEE 802.11ay \cite{wu2019fast}. Furthermore, different transmission times of two packets in a directional FD link can affect the channel utilization and reduce achievable link throughput. 
\end{itemize}

In this paper, we introduce DFDMAC, a directional FD MAC protocol, to address the challenges mentioned above and design a power control algorithm to boost the performance of mmWave FD networks. Firstly, DFDMAC extends the RTS/CTS handshaking in IEEE 802.11ay to enable two-node and three-node directional FD transmissions. Specifically, we redesign the frame structures of RTS and CTS to convey information about a node's duplex mode and work mode. FD is also used to exchange control frames for reducing overhead in the three-node directional FD transmission. In addition, to prevent deafness and directional HN problems in mmWave FD networks, DFDMAC features a novel busy-tone mechanism that leverages omnidirectional transmission of out-of-band signals to improve channel access fairness. Secondly, to minimize the impact of residual SI on the FD link throughput, we formulate the FD link throughput optimization as a channel occupation time minimization problem for the two typical FD transmission modes and propose a power control algorithm to determine the optimal transmit power. Extensive simulation results demonstrate that DFDMAC can remarkably improve the network throughput and channel access fairness performance by over $60\%$ and $32\%$, compared with the MAC protocol in the state-of-the-art IEEE 802.11ay. Furthermore, our proposed power control algorithm effectively enhances network throughput by matching higher transmission rates. The main contributions of this paper are summarized as follows:
 \begin{itemize}
 	\item We propose a distributed directional FD MAC protocol that supports typical directional FD transmission modes;
 	\item We design a novel busy-tone mechanism that overcomes deafness and directional HN problems in mmWave FD networks;
 	\item We design a power control algorithm that enhances channel utilization and FD link throughput.
 \end{itemize}

The remainder of this paper is organized as follows. Section~\ref{protocol} describes the details of the proposed DFDMAC protocol. Section~\ref{model} presents the system model and throughput optimization based on power control. Simulation results are presented in Section~\ref{simulation}, and the paper is concluded in Section~\ref{conclusion}.

\section{Directional Full-Duplex MAC Protocol} \label{protocol}

This section first introduces the network scenario considered in this paper. Then, the key components in the proposed MAC protocol, i.e., newly designed frame structures of RTS and CTS, a novel busy-tone mechanism, and details on establishing two typical directional FD transmissions are presented. Finally, we point out asymmetric FD transmission issue which can cause the FD link throughput degradation. 

\subsection{mmWave FD Network}
\begin{figure}[t]
	\centering
	\includegraphics[height=4cm]{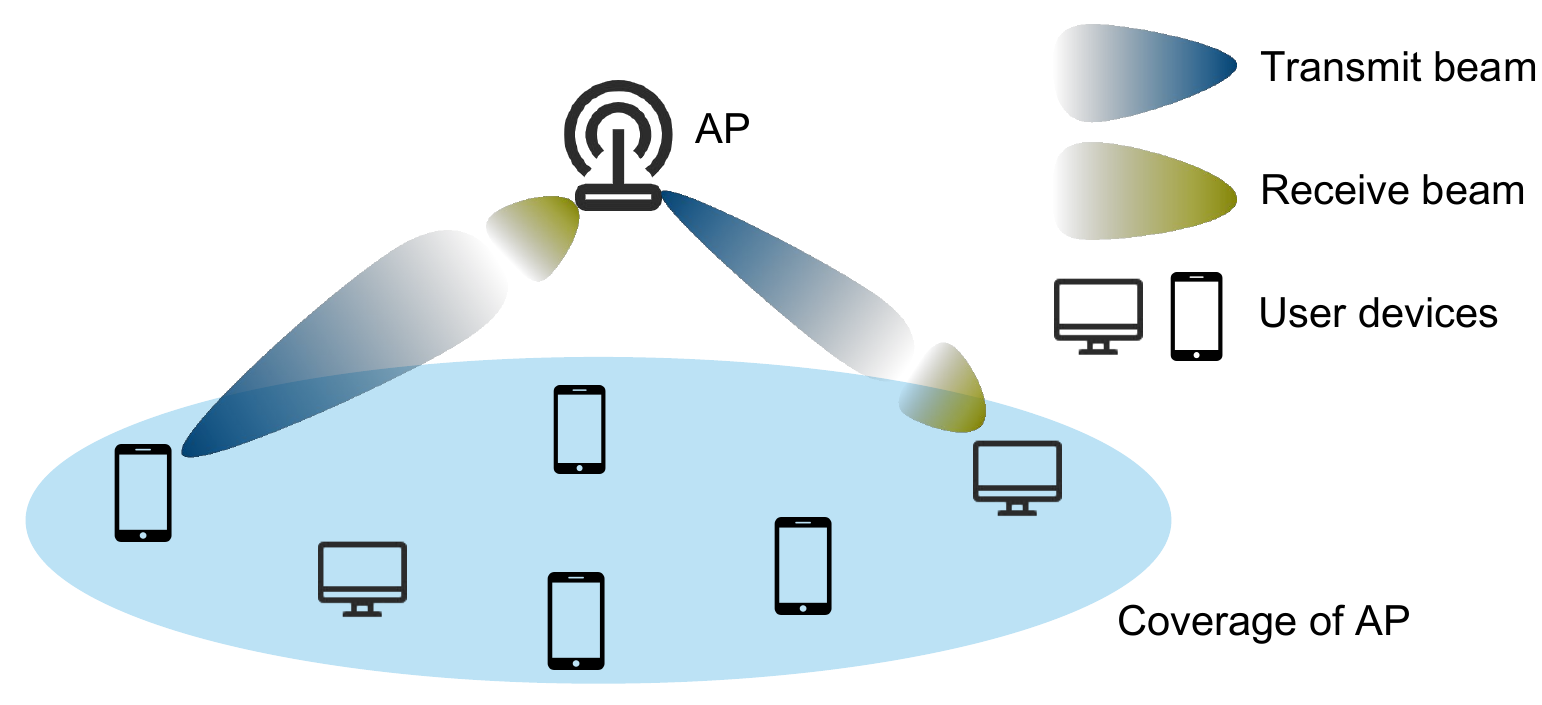}
	\caption{Considered mmWave FD network.}\label{fig_network}
\end{figure}

We consider a distributed mmWave FD network comprising an access point (AP) and multiple user devices. As shown in Fig. \ref{fig_network}, each node supports FD communication and directional transmission and reception using its transmit and receive beams. With FD capability, a node can transmit a packet while simultaneously receiving from another node. As a result, FD transmissions in the mmWave FD network are categorized into two distinct modes. 
\begin{figure}[t]
	\centering
	\includegraphics[scale=0.58]{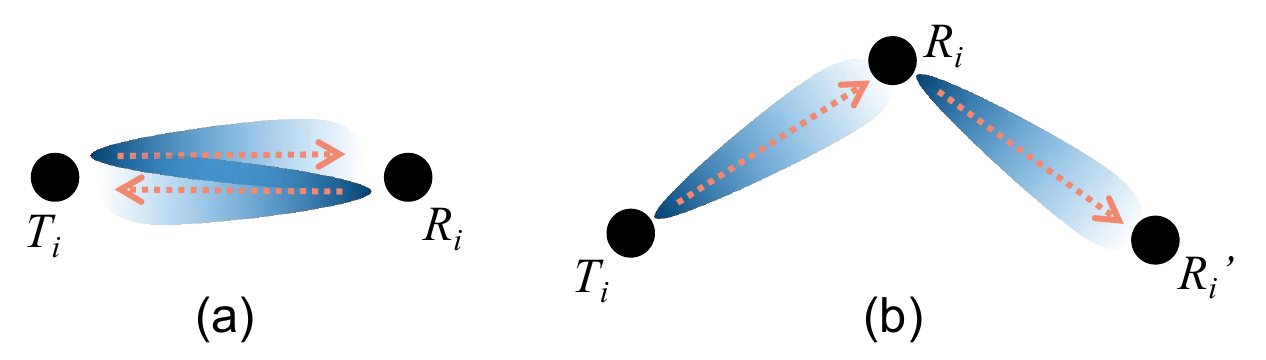}
	\caption{Two typical directional FD transmission modes: (a) Two-node directional FD transmission; (b) Three-node directional FD transmission.}\label{fig_modes}
\end{figure}


\begin{itemize}
	\item Two-node directional FD transmission mode: As shown in Fig. \ref{fig_modes} (a), transmitter $T_i$ wins the channel and initiates a \textit{\textbf{primary}} directional transmission to receiver $R_i$. At the same time, \rev{$R_i$ enables a \textit{\textbf{secondary}} transmission to $T_i$}; 
	\item Three-node directional FD transmission mode: As shown in Fig. \ref{fig_modes} (b), transmitter $T_i$ wins the channel and initiates a \textit{\textbf{primary}} directional transmission to the primary receiver $R_i$. At the same time, $R_i$ enables a \textit{\textbf{secondary}} transmission to secondary receiver $R_i'$. 
\end{itemize}


\subsection{Frame Structure Design}
\begin{figure}[t]
	\centering
	\includegraphics[scale=0.45]{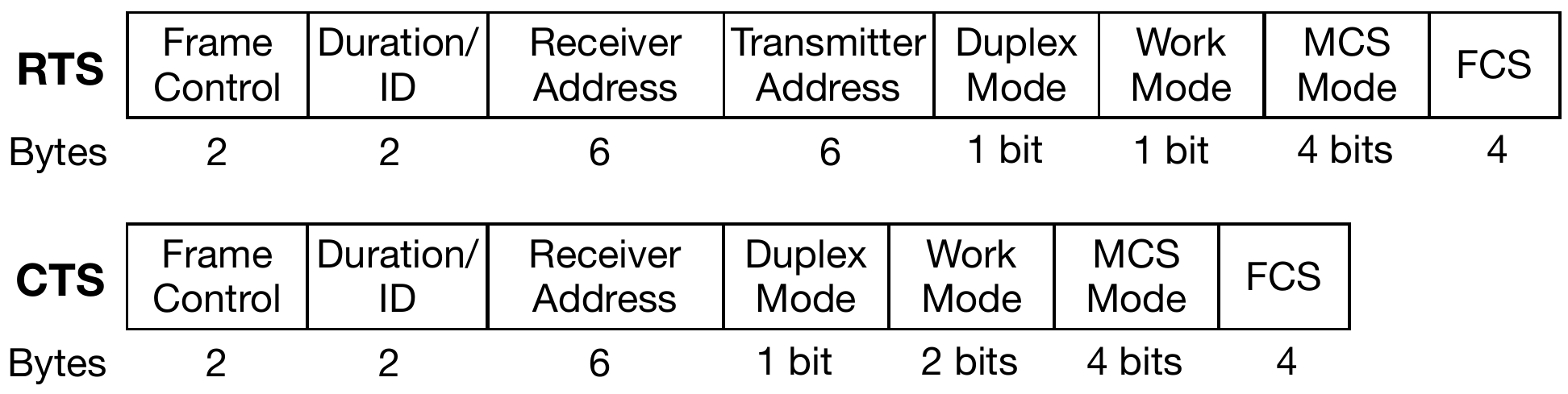}	
	\caption{\rev{Frame structures of RTS and CTS.}}\label{fig_frame}
\end{figure}

The DFDMAC uses the exchange of RTS and CST frames to establish the two typical directional FD transmission modes, which is not supported in the existing RTS/CTS handshaking in IEEE 802.11ay. Thus, we redesign the frame structures of RTS and CTS to convey important information, including duplex mode, work mode, and \rev{modulation and coding
	schemes (MCS)} mode, as shown in Fig.~\ref{fig_frame}. 

\textbf{RTS frame}: In the DFDMAC protocol, the RTS frame is a request signal from a transmitter to a receiver. To support the mentioned directional FD transmissions, we add three new fields, i.e., duplex mode, work mode, and MCS mode. The duplex mode field contains $1$ bit and indicates if the transmitter supports FD, where $0$ represents HD and $1$ represents FD. The MCS mode field contains $4$ bits and indicates the physical transmission rate of the DATA frame. The work mode field contains \rev{$1$} bits and describes two cases:
\begin{itemize}
	\item \rev{$Work\enspace Mode=0$}: The RTS frame is transmitted by a primary transmitter without indicating the transmission mode, which is determined by the primary receiver;
	\item \rev{$Work\enspace Mode=1$}: The RTS frame is transmitted by a secondary transmitter and used to inform the receiver to work in HD receiving mode.
\end{itemize}


\textbf{CTS frame}: In the DFDMAC protocol, the CTS frame is a response signal from the receiver to the transmitter. Similar to the RTS frame, we also add the three new fields. The duplex mode field contains $1$ bit and indicates if the receiver supports FD. The MCS mode field contains $4$ bits and indicates the physical transmission rate of the DATA frame. The work mode field contains $2$ bits, indicating the following three cases:
\begin{itemize}
	\item $Work\enspace Mode=00$: The node transmitting the CTS frame will work in HD mode to receive a DATA frame;
	\item $Work\enspace Mode=01$: The node transmitting the CTS frame will work in two-node FD mode to transmit and receive a DATA frame simultaneously;
	\item $Work\enspace Mode=10$: The node transmitting the CTS frame will work in three-node FD mode to transmit a DATA frame to another node while receiving.
\end{itemize}

\subsection{Busy-Tone Mechanism} \label{tone}

In the DFDMAC protocol, we design a novel busy-tone (BT) mechanism to avoid deafness and HN problems in mmWave FD networks. Unlike existing BT mechanisms that only consider a transmitter and cannot handle situations where there are two transmitters in an FD link \cite{BT_2020}, our proposed BT mechanism can prevent both a transmitter and its receiver from becoming deaf nodes. The proposed BT mechanism has the following characteristics.\footnote{IEEE 802.11ay supports fast session transfer protocol which makes it backward compatible with 2.4 GHz or 5 GHz WLAN \cite{80211ay}. Thus, a mmWave node can omnidirectionally transmit a BT signal with some unused frequency in unlicensed 2.4 GHz or 5 GHz bands, which can cover a much large area than directional mmWave transmission with the same transmit power.}
\begin{itemize}
	\item The BT signal is an out-of-band sine-wave signal with a unique frequency and is transmitted omnidirectionally;
	\item The duration of a BT signal is very short, less than 1 slot;
	\item Each node has a \textit{start BT} and  an \textit{end BT} indicating the beginning and end of a transmission, respectively.
\end{itemize}

In the BT mechanism, each node keeps detecting the BT signal while listening to the mmWave channel omnidirectionally as that in IEEE 802.11ay. When a node wins the channel, it transmits its own, and its receiver's start BT signals while starting to transmit the RTS frame. Once a node overhears its start BT signal, it immediately transmits its start BT signal again as a response. If a node overhears its intended receiver's start BT signal when executing the backoff mechanism to contend the channel, the node will freeze its backoff counter and defer its transmission until receiving its receiver's end BT signal. More details about the BT mechanism are introduced in the next subsection.

Traditional omnidirectional carrier-sensing mechanisms cannot provide accurate channel state information in mmWave FD networks with directional transmissions, which leads to deafness and HN problems that degrade network performance. The proposed BT mechanism effectively overcomes these issues and offers the following benefits:
\begin{itemize}
	\item Avoiding deafness problem. A node $N_i$'s omnidirectional BT signal can be received by neighboring nodes when they fail to receive $N_i$'s directional RTS or CTS frame, thus avoiding unnecessarily increasing the contention window;
	\item Avoiding directional HN problem. Neighboring nodes with packets to send will stop transmitting when they cannot receive directional RTS or CTS frames but overhear omnidirectional BT signals from the intended nodes, preventing the HN problem.
\end{itemize}

\subsection{Operation of DFDMAC}
With the control frames and the BT mechanism, two-node and three-node directional FD transmissions can be established. Next, we describe the procedure of establishing the two types of FD transmissions in detail.

\subsubsection{Two-Node Directional FD Transmission}
\begin{figure}[t]
	\centering
	\includegraphics[scale=0.4]{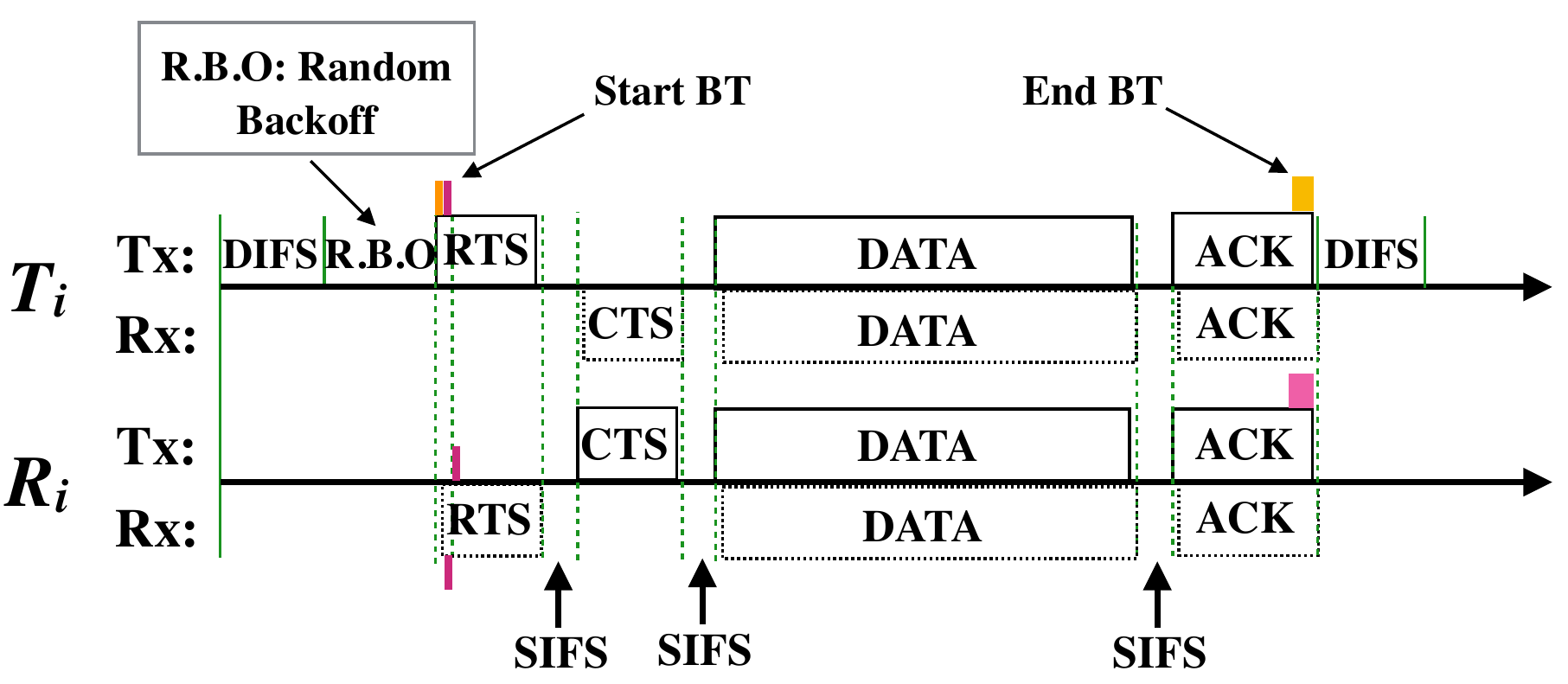}
	\caption{Illustrative example of a two-node directional FD transmission.}\label{fig_MAC2}
\end{figure}

Figure \ref{fig_MAC2} shows a two-node directional FD transmission between node $T_i$ and $R_i$. The initial transmitter $T_i$ wins the channel with the random backoff mechanism and then initiates a transmission to $R_i$. Specifically, $T_i$ omnidirectionally transmits its start BT signal and its receiver's start BT signal, while starting to transmit an RTS frame to $R_i$ using a transmit beam. $R_i$ transmits its start BT once detecting it. After receiving the RTS frame in omnidirectional mode, $R_i$ waits for a SIFS time and directionally transmits a CTS frame to $T_i$. If $R_i$ has a packet for $T_i$ and the work mode field in the received RTS frame is $00$, the work mode field in the CTS frame is set to $01$. After transmitting the RTS frame, $T_i$ waits to receive the CTS frame with its receiving beam pointing to $R_i$. If the value of work mode in the received CTS frame is $01$, $T_i$ knows that $R_i$ also has a packet for $T_i$. Then $T_i$ waits for a SIFS time and then gets prepared to receive the DATA frame from $R_i$ with its receiving beam while directionally transmitting its DATA frame to $R_i$. After receiving the DATA frames, $T_i$ and $R_i$ wait a SIFS time and then directionally transmit ACK frames simultaneously. At the end of transmitting the ACK frames, both $T_i$ and $R_i$ transmit their ending BT signal omnidirectionally.

\subsubsection{Three-Node Directional FD Transmission}
\begin{figure}[t]
	\centering
	\includegraphics[scale=0.4]{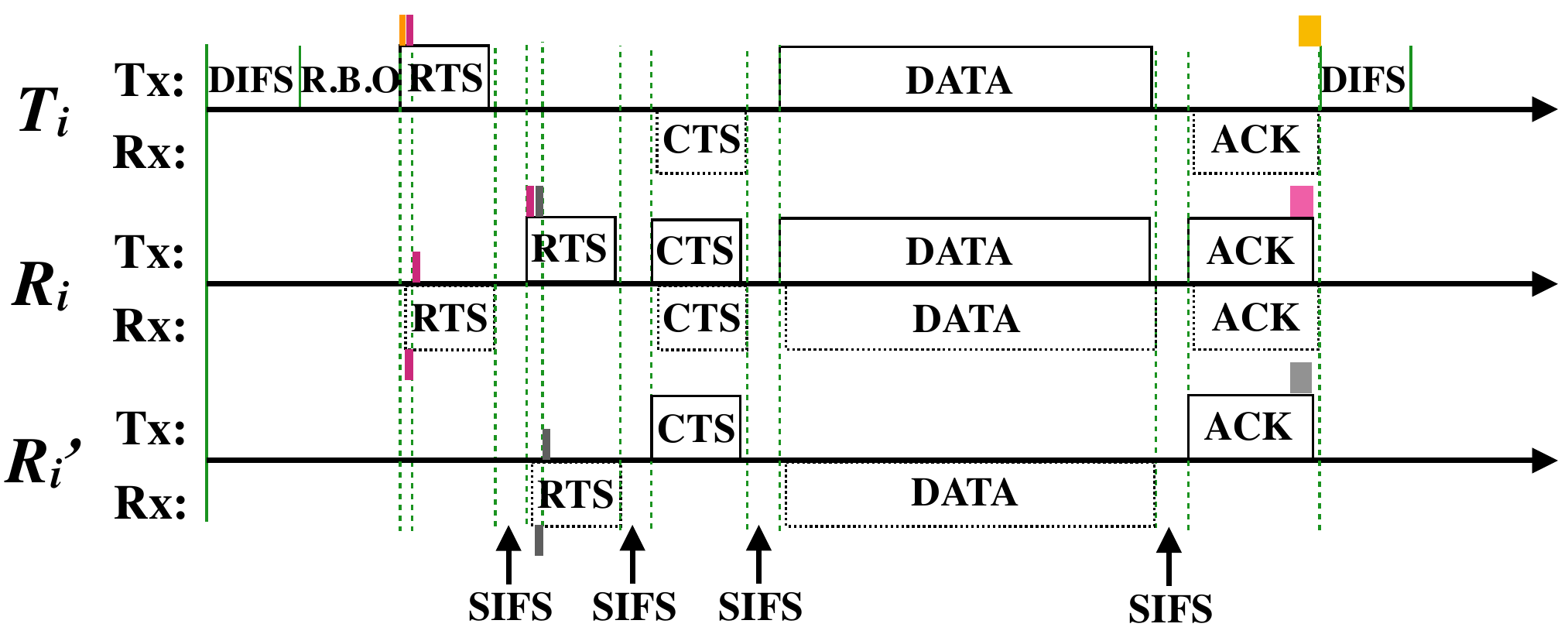}
	\caption{Illustrative example of a three-node directional FD transmission.}\label{fig_MAC3}
\end{figure}
Figure \ref{fig_MAC3} shows a three-node directional FD transmission. Specifically, the primary transmitter $T_i$ wins the channel and transmits an RTS frame to the primary receiver $R_i$, meanwhile omnidirectionally transmitting its start BT signal and its receiver's start BT signal. $R_i$ transmits its start BT once detecting the signal. After receiving the primary RTS frame, if $R_i$ has a packet for another node $R_i'$ and $T_i$'s transmission does not interfere with the reception of $R_i'$, it waits a SIFS time and directionally transmits its RTS frame with the work mode set to $10$ to the secondary receiver $R_i'$. Meanwhile, $R_i$ also transmits its start BT signal and its receiver's start BT signal. $R_i'$ transmits its start BT once detecting the signal. After receiving the secondary RTS frame, $R_i'$ waits a SIFS time and directionally transmits a CTS frame to $R_i$. At the same time, $R_i$ also directionally transmits its CTS frame to $T_i$. After directionally receiving the CTS frames, $T_i$ and $R_i$ directionally transmit their DATA frames to $R_i$ and $R_i'$, respectively. After receiving the DATA frames, $R_i$ and $R_i'$ simultaneously transmit ACK frames to $T_i$ and $R_i$, respectively. At the end of transmitting ACK frames, the three nodes transmit their end BT signals omnidirectionally.

It is worth noting that, the overhead caused by the transmissions of control frames plays an important role in the network throughput because the transmission rate of the control frame is much lower than that of the DATA frame. In order to reduce such overhead, we exploit FD to transmit ACK frames simultaneously in the proposed MAC protocol. Furthermore, a three-node FD transmission only needs an extra RTS frame and a SIFS by transmitting two CTS frames simultaneously, compared with a two-node FD transmission.

\subsection{Asymmetric Transmission Issue}\label{protocol_issue}
\begin{figure}[t]
	\graphicspath{{figure/}}
	\centering
	\subfigure[]{
		\includegraphics[height=2.5cm]{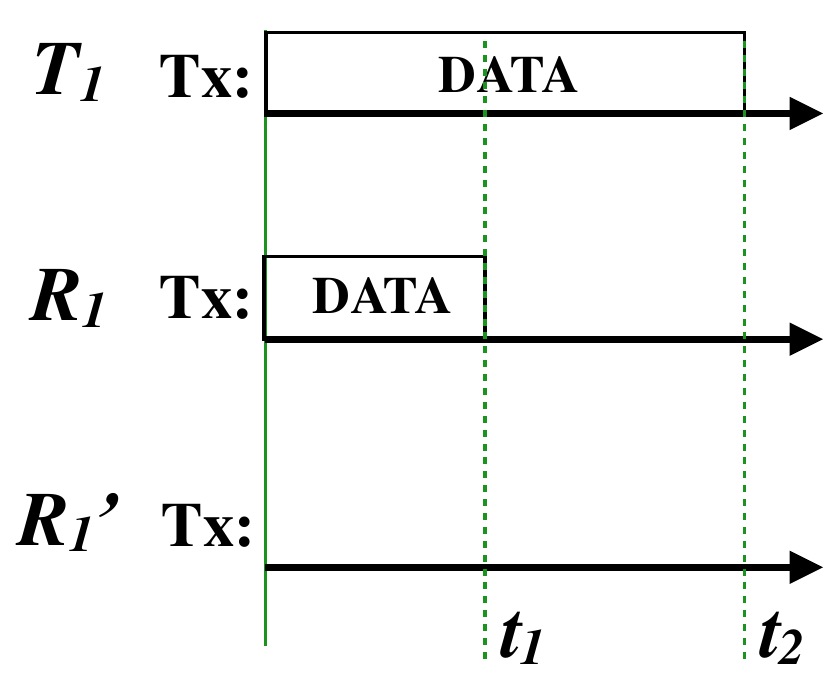}
	}
	\subfigure[]{
		\includegraphics[height=2.5cm]{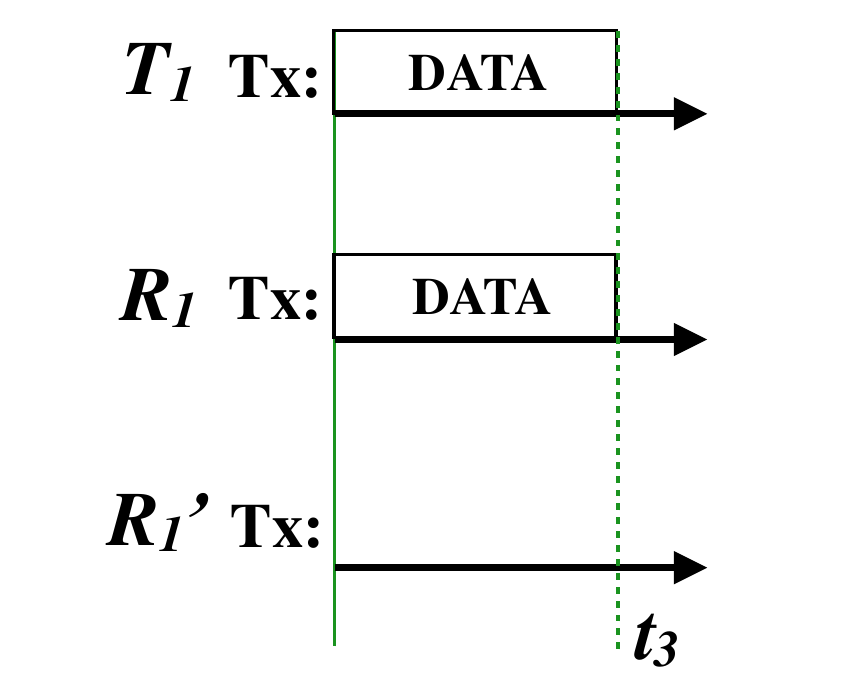}
	}
	\caption{Illustrative examples: (a) asymmetric FD transmission; (b) symmetric FD transmission.} \label{fig_problem}
\end{figure}

For an FD transmission link in practical mmWave FD networks, the transmission time of the primary transmitter's DATA frame can be different from that of the secondary transmitter's  DATA frame, due to varying payload size and different transmission rates affected by the transmit power, channel condition, SI cancellation level, etc. In this case, the channel occupation time of the FD link is determined by the longer DATA frame's transmission time. This issue reduces the channel utilization and results in FD link throughput degradation. In order to solve the issue and enhance the FD link throughput, we can adjust the transmit powers of the two transmitters in an FD link to optimize the received SINRs and transmission rates, which can reduce the channel occupation time. For instance, as shown in Fig. \ref{fig_problem}, $T_1$ and $R_1$ in an FD link need $t_1$ and $t_2$ ($t_1 < t_2$) to finish the DATA transmission, respectively. Thus, the channel occupation time of the FD link is $t_2$. However, with power control and transmission rate matching, the FD link only needs $t_3$ to transmit the two DATA frames, where $t_1 < t_3 < t_2$. Therefore, to maximize the channel utilization and FD link throughput, an efficient power control algorithm is desired to obtain the optimal transmit power.

\section{System Model and Throughput Optimization} \label{model}
\subsection{Directional Antenna Model}

Consider a widely used directional antenna model\cite{Interference_ToN11}. The antenna of a node has $M$ beams that can cover all directions, where $M=2\pi / \theta_{N_i}$, and $\theta_{N_i}$ denotes the beamwidth in radians of node $N_i$.
When node $T_i$ transmits a signal to receiver $R_i$ using a transmit beam, the transmit antenna gain at $T_i$ is given by
$G_{T_i}^{Tx} = g(\phi_t)G_{T_i}^{max}$,
where $G_{T_i}^{max}$ denotes the maximum transmit antenna gain of $T_i$\footnote{The maximum antenna gain is closely related to the number of antenna arrays, angle of arrival and angle of incidence of beams\cite{Pielli}. In this paper, $G_{N_i}^{max}=\frac{2\pi}{\theta_{N_i}}$.}. Here,  $g(\phi_t)$ is given by 
\begin{equation}
	g(\phi_t) = \left\{
	\begin{matrix}
		1, &|\phi_t|<\frac{\theta_{T_i}}{2}  \\
		0, & \text{otherwise}
	\end{matrix}
	\right.,
\end{equation}
where $\phi_t$ denotes the relative angle with respect to the boresight of the transmit beam. Note that $g(\phi_t)$ is used to determine if $R_i$ is located in the coverage of $T_i$'s transmit beam.

When $R_i$ receives a signal from $T_i$ with a receive beam, the receive antenna gain at $R_i$ is given by
$G_{R_i}^{Rx} = g(\phi_r)G_{R_i}^{max}$,
where $G_{R_i}^{max}$ denotes the maximum receive antenna gain of $R_i$. Here, $g(\phi_r)$ is given by 
\begin{equation}
	g(\phi_r) = \left\{
	\begin{matrix}
		1, &|\phi_r|<\frac{\theta_{R_i}}{2} \\
		0, & \text{otherwise}
	\end{matrix}
	\right.,
\end{equation}
where $\phi_r$ denotes the relative angle with respect to the boresight of the receive beam. 

\subsection{Transmission Model}

We adopt the Friis transmission model for mmWave signal propagation. Then, the path gain from $T_i$ to $R_i$ is given by
\begin{equation}\label{pathloss}
G(T_i, R_i) = G_0d(T_i, R_i)^{-\alpha}e^{-c_o d(T_i, R_i)},
\end{equation}
where $d(T_i, R_i)$ is the distance between $T_i$ and $R_i$, $G_0$ denotes the path loss gain of mmWave signal at the reference distance of $1$ m, $\alpha$ denotes the path-loss exponent, and $c_0$ denotes the attenuation factor due to the oxygen absorption loss ( $c_0 = 0.0037$/m in \cite{Interference_ToN11}).  
The received signal strength is given by
\begin{equation} \label{rss}
P(T_i, R_i) = P_{T_i} G_{T_i}^{Tx} G_{R_i}^{Rx} G(T_i, R_i),
\end{equation}
where $P_{T_i}$ is the transmit power.

In the mmWave FD network, we assume that a node can successfully decode the received signal only if the signal-to-interference-plus-noise ratio (SINR) exceeds a given threshold. 
Next, we analyze the successful SINR conditions separately for two-node and three-node FD transmissions.
\subsubsection{Two-Node Directional FD Transmission}
For a two-node FD link, $T_i$ and $R_i$ receive DATA frames from each other. According to equation \eqref{rss}, the received SINRs of $T_i$ and $R_i$ can be respectively expressed as
\begin{equation}\label{eq_2SINR2}
	SINR_{T_i} = \frac{P_{R_i} G_{R_i}^{Tx} G_{T_i}^{Rx} G(R_i, T_i)}{I^{SI}_{T_i}+I_s+n_0},
\end{equation}
\begin{equation}\label{eq_2SINR1}
	SINR_{R_i} = \frac{P_{T_i} G_{T_i}^{Tx} G_{R_i}^{Rx} G(T_i, R_i)}{I^{SI}_{R_i}+I_s+n_0},
\end{equation}
where $P_{R_i}$ is the transmit power of $R_i$, $I^{SI}_{T_i}$ and $I^{SI}_{R_i}$ denote the residual SI at $T_i$ and $R_i$, respectively, $n_0$ is the background noise, and $I_s$ denotes the interference from nearby concurrent links. 
Then, the SINR thresholds for successful transmission from $R_i$ to $T_i$ and from $T_i$ to $R_i$ are given by
$SINR_{T_i}  \geq \gamma_{T_i}, 
SINR_{R_i} \geq \gamma_{R_i},$
which ensure that $T_i$ and $R_i$ successfully receive the intended DATA frames. 
Let $\beta_{N_i}$ be the SI cancellation level at an FD node $N_i$. Then, the residual SI at $N_i$ is given by
\begin{equation}\label{eq_SI}
I^{SI}_{N_i} = P_{N_i} \beta_{N_i}.
\end{equation}

\subsubsection{Three-Node Directional FD Transmission}
For three-node directional FD link pair, $T_i$ and $R_i$ transmit DATA frames to $R_i$ and $R_i'$, respectively. Similarly, the received SINRs of $R_i$ and $R_i'$ can be expressed as
\begin{equation}\label{eq_3SINR1}
	SINR_{R_i} = \frac{P_{T_i} G_{T_i}^{Tx} G_{R_i}^{Rx} G(T_i, R_i)}{I^{SI}_{R_i}+I_s+n_0},
\end{equation}
\begin{equation}\label{eq_3SINR2}
	SINR_{R_i'} = \frac{P_{R_i} G_{R_i}^{Tx} G_{R_i'}^{Rx} G(R_i, R_i')}{P_{T_i} G_{T_i}^{Tx} G_{R_i'}^{Rx} G(T_i, R_i')+I_s+n_0}.
\end{equation}
Then, the SINR thresholds that guarantee successful transmission from $T_i$ to $R_i$ and from $R_i$ to $R_i'$ are given by
$SINR_{R_i} \geq \gamma_{R_i},
SINR_{R_i'}  \geq \gamma_{R_i'}.$
Note that $T_i$'s transmission can cause inter-beam interference (IBI) to $R_i'$ in a three-node directional FD link, when $T_i$'s transmit beam is directed at $R_i'$'s receive beam. 

Furthermore, the equations \eqref{eq_SI} and \eqref{eq_3SINR2} show that the transmit power has a crucial impact on the residual SI and IBI, which can further affect the received SINR. Therefore, we should adjust the transmit powers of the primary and the secondary transmitters in an FD link to effectively improve the link throughput via optimizing the achieved SINRs.

\subsection{Throughput Optimization}\label{Optimization}
In order to maximize the FD link throughput, the asymmetric transmission issue as mentioned in Section \ref{protocol_issue} needs to be properly handled. To this end, we first define the achievable throughput of an FD link, which is given by 
	\begin{equation}\label{eq_S}
		S = \frac{L_{T_i}+L_{R_i}}{T_{overhead}+\max \left(\frac{L_{T_i}}{r_{T_i}},\frac{L_{R_i}}{r_{R_i}}\right)},
	\end{equation}
where $L_{T_i}$ and $L_{R_i}$ denote the given payload size in $T_i$'s and $R_i$'s DATA frames, respectively, $T_{overhead}$ denotes the total time of transmitting control frames, transmitting the header of DATA frame from the physical layer, and the inter-frame spacing at the MAC layer, $r_{T_i}$ and $r_{R_i}$ denote the physical transmission rates of $T_i$'s and $R_i$, respectively. It is worth noting that the achievable throughput defined in \eqref{eq_S} is applied to both two-node and three-node FD transmission links because the primary and the secondary transmitters are the same in both types of FD links, i.e., $T_i$ and $R_i$.

In a practical wireless network, only a given number of discrete physical transmission rates can be used in the IEEE 802.11ay standard. A certain transmission rate can be supported only if the achieved SINR is larger than a corresponding SINR threshold. Thus, $r_{T_i}$ and $r_{R_i}$ are the highest rates that can be supported by the achieved SINRs at the primary and secondary receivers. Then, we have $r_{T_i}=f(SINR_{R_i})$ and $r_{R_i}=f(SINR_{T_i})$ for the two-node directional FD transmission link, and $r_{T_i}=f(SINR_{R_i})$ and $r_{R_i}=f(SINR_{R_i'})$ for the three-node directional FD transmission link. According to \eqref{eq_2SINR2}, \eqref{eq_2SINR1}, \eqref{eq_3SINR1}, and \eqref{eq_3SINR2}, the achieved SINRs are determined by the transmit powers of $T_i$'s and $R_i$. Therefore, adjusting the transmit power can optimize the achieved SINR to maximize the FD link throughput.

Since $T_i$ and $R_i$ transmit their DATA frames simultaneously, maximizing the FD link throughput $S$ is equivalent to minimizing the channel occupation time, which is given by 
\begin{equation}\label{eq_D}
	D = \max \left(\frac{L_{T_i}}{r_{T_i}},\frac{L_{R_i}}{r_{R_i}}\right ).
\end{equation}
Then, the throughput optimization problem boils down to minimizing the channel occupation time, which is given by
\begin{equation}\label{P_1}
	\begin{aligned}
			& \text{\textbf{P1}:}
			& & \min_{P_{T_i}^{Tx}, P_{R_i}^{Tx}} D \\
			& \mathrm{s.t.} 
			& & P^{min}_{T_i}<P_{T_i}<P^{max}_{T_i}, \\
			& \mathrm{\quad}
			& & P^{min}_{R_i}<P_{R_i}<P^{max}_{R_i}.
		\end{aligned}
\end{equation}
where $P_{N_i}^{min}$ and $P_{N_i}^{max}$ denote the minimum and maximum transmit power of node $N_i$.

The transmit power in practical communication systems is discrete, such as the integers in the range $[- 12, 19]$~dBm with a minimum interval $\Delta$ \cite{pocmac}. 
Hence, possible combinations of the transmit powers  $(P_{T_i}^{}, P_{R_i}^{})$ are limited. To obtain the optimal transmit powers $(P_{T_i}^{*}, P_{R_i}^{*})$, we design a power control algorithm shown in Alg. \ref{alg}. In the algorithm, the primary transmitter $T_i$ first uses the maximum transmit power. Then, the minimum and the maximum transmission rates of $T_i$ are derived by SINR calculation and rate matching when the secondary transmitter $R_i$ uses the maximum and the minimum transmit power, respectively. For each transmission rate $r_{T_i}$ supported by $T_i$, the required SINR can be obtained, which can be used to derive the transmit power $P_{R_i}$. With $P_{T_i}$ and $P_{R_i}$, the achieved SINR of $P_{T_i}$ can be caculated to derive the transmission rate $r_{R_i}$. Using $r_{T_i}$ and $r_{R_i}$, the FD link throughput can be calculated to update the optimal transmit powers ($P_{T_i}^*$, $P_{R_i}^*$). The optimal transmit powers ($P_{T_i}^*$, $P_{R_i}^*$) are obtained until $P_{T_i}$ is smaller than the minimum transmit power. 

\begin{algorithm}[t]\small
	\caption{Power Control Algorithm }\label{alg}
	\begin{algorithmic}[1]
		\STATE {$P_{T_i} = P_{T_i}^{max}$;}
		\WHILE{$P_{T_i} \geq P_{T_i}^{min}$}	
		\STATE {Derive $r_{T_i}^{min}$ ($P_{R_i} = P_{R_i}^{max}$) and $r_{T_i}^{max}$  ($P_{R_i} = P_{R_i}^{min}$);}	 
		\FOR{$r_{T_i}$ in $[r_{T_i}^{min}, r_{T_i}^{max}]$}
		\STATE Derive $P_{R_i}$ with $r_{T_i}$ and $P_{T_i}$;
		\STATE Derive $r_{R_i}$ with $P_{R_i}$ and $P_{T_i}$;
		\IF{$D < max(\frac{L_{R_i}}{r_{R_i}}, \frac{L_{T_i}}{r_{T_i}})$}
		\STATE $D = max(\frac{L_{R_i}}{r_{R_i}}, \frac{L_{T_i}}{r_{T_i}})$;
		\STATE Update the optimal transmit powers $(P_{T_i}^{*},P_{R_i}^{*})$;
		\ENDIF
		\ENDFOR
		\STATE $P_{T_i} = P_{T_i} - \Delta$;
		\ENDWHILE
		\RETURN  $(P_{T_i}^{*},P_{R_i}^{*})$;
	\end{algorithmic}
\end{algorithm}


\section{Simulation Results}\label{simulation}

 \begin{table}[t]
	\centering\caption{Simulation Parameters}
	\label{para}
	\begin{tabular}{ | l |  c | l  | c | }
		\hline
		\textbf{Parameter}   & \textbf{Value}         & \textbf{Parameter}   & \textbf{Value}\\\hline
Control PHY header     & $40$ bits  &  DIFS  & $13$ us \\\hline
SC PHY header         & $64$ bits & SIFS      & $3$ us \\\hline
MAC header     & $320$ bits &  Slot time  & $5$ us\\\hline
Packet payload      & $8000$ bytes &  $CW_{min}$ & $ 16$ \\ \hline
Control PHY rate & $ 27.5$ Mbps  &  $CW_{max}$      & $1024$ \\\hline
RTS         & $352$ bits & $\beta$     & $-85$ dB\\\hline
CTS        & $304$ bits & $n_0$       & $-90$ dBm\\\hline
ACK       & $304$ bits  &  $\alpha$  & $2$ \\ \hline
	\end{tabular}
\end{table}


 \begin{table}[t]
	\centering\caption{MCS and the Corresponding SINR Thresholds}
	\label{MCS}
	\begin{tabular}{ | l | c | c | c | }
		\hline
		MCS   	 & MCS 1   & MCS 2   & MCS 3\\\hline
		Modulation   & QPSK    & QPSK  & 16-QAM \\\hline
		Coding rate    & $1/2$  & $2/3$    & $2/3$ \\\hline
		Data rate     & $952$ Mbps &  $1904$ Mbps  & $3807$ Mbps\\\hline
		SINR threshold & $5.5$ dB &  $13$ dB & $18$ dB \\ \hline
	\end{tabular}
\end{table}

\subsection{Performance Evaluation of the proposed DFDMAC}
\begin{figure}[t]
	\graphicspath{{figure/}}
	\centering
	\subfigure[Saturation throughput]{
		\includegraphics[width=4.1cm]{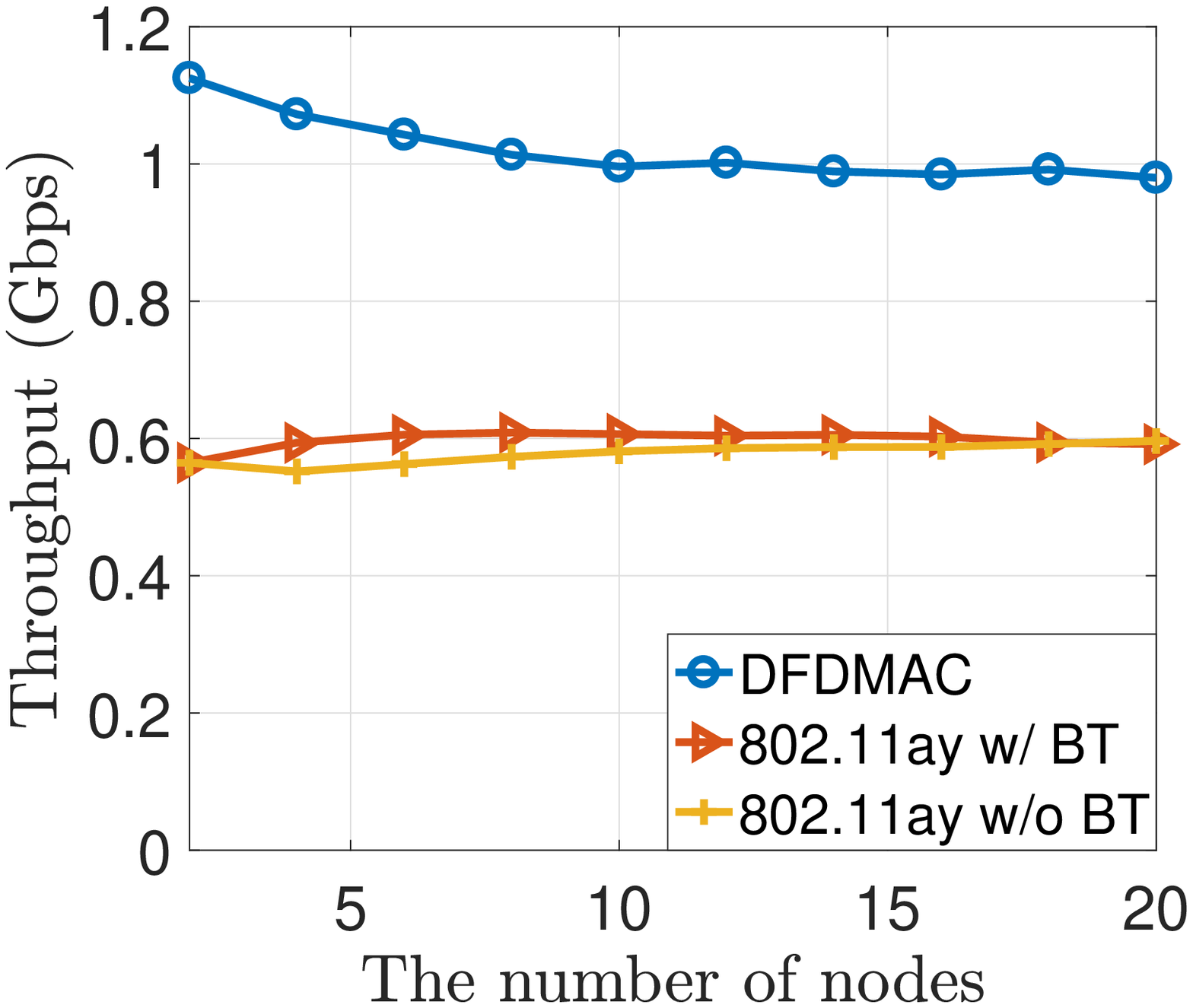}
	}
	\subfigure[Fairness performance]{
		\includegraphics[width=4.1cm]{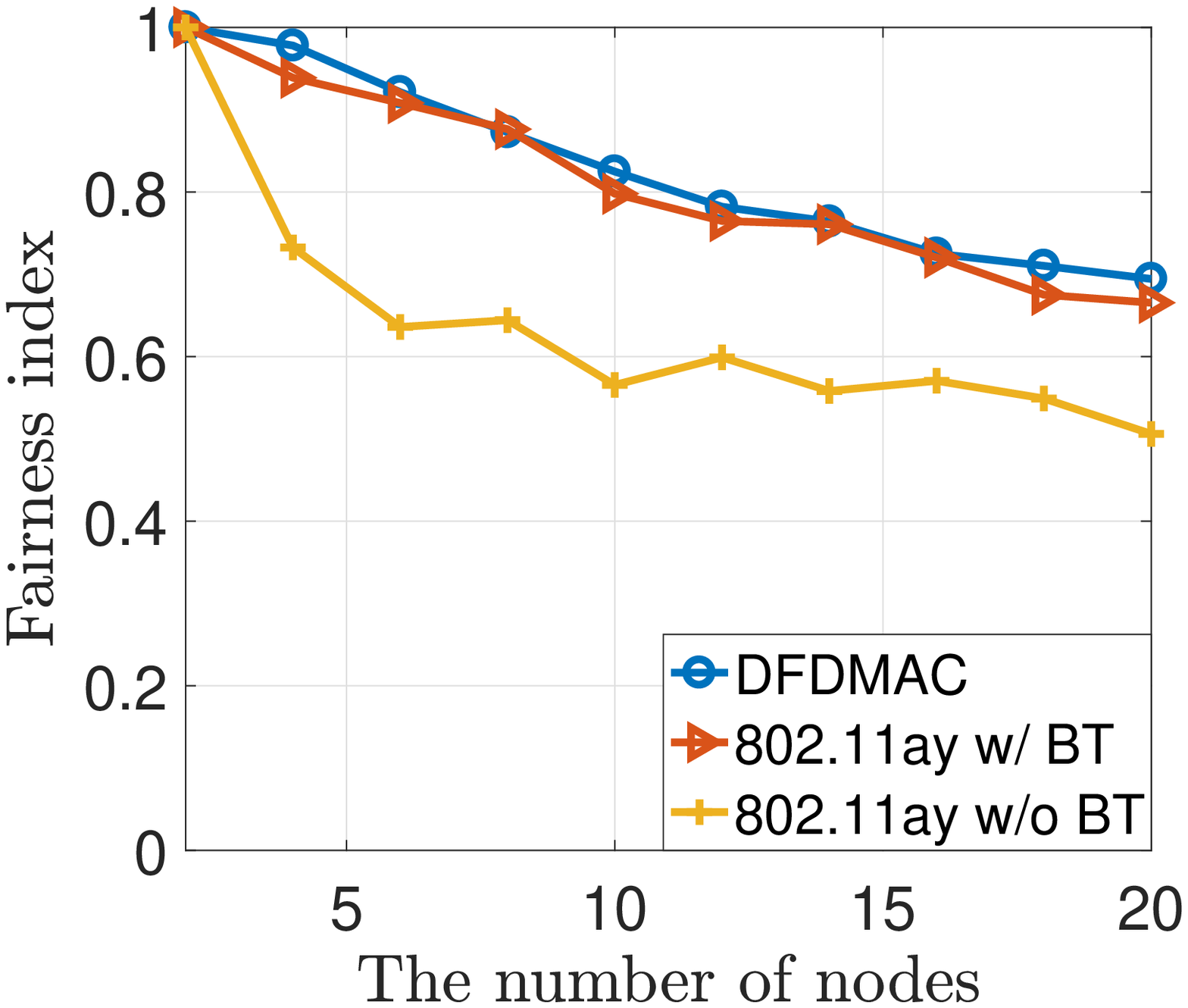}
	}
	\caption{Performance of the proposed DFDMAC protocol in terms of saturation throughput and accessing fairness.} \label{fig_HDvsFD}
\end{figure}

We \rev{use the discrete event simulator provided in \cite{DMAC_11ad} to} evaluate the performance of the DFDMAC in terms of throughput and fairness. We simulate a mmWave FD network with $n$ nodes, including an AP and $n-1$ user devices, which are randomly distributed in a circular area with a radius of $10$ m. We consider a saturated traffic scenario. All the user devices always have packets for the AP, while the AP randomly selects a user device to transmit a packet after successfully finishing a transmission. Each node has $12$ beams and the optimal transmit and receive beam pairs are identified beforehand. The transmission rate is $1904$ Mbps. Important simulation parameters are listed in Table \ref{para}. Regarding the benchmarks, the proposed DFDMAC is compared with the following two MAC protocols:
\begin{itemize}
	\item \textbf{802.11ay w/ BT}: IEEE 802.11ay protocol with the proposed BT mechanism;
	\item \textbf{802.11ay w/o BT}: IEEE 802.11ay protocol without the proposed BT mechanism.
\end{itemize}

Figure \ref{fig_HDvsFD} (a) shows the network throughput performance of all the three MAC protocols with respect to the number of nodes. From the figure, we can observe that the DFDMAC can achieve the highest network throughput, which is $60\%$ higher than that achieved by IEEE 802.11ay. Due to the overhead caused by the transmission of control frames and random backoff mechanism, the FD gain in throughput in the mmWave network cannot reach $100\%$. In addition, IEEE 802.11ay with the proposed BT mechanism cannot achieve better throughput performance compared with IEEE 802.11ay. It illustrates that the HN problem is not severe, and the channel utilization is not affected by the deafness problem in HD mmWave networks. 

Since deafness problem can result in unfair channel access, we adopt Jain's fairness index defined in \cite{fair} to evaluate the fairness performance among all user devices.
Fig. \ref{fig_HDvsFD} (b) shows the throughput fairness with respect to the number of nodes. It can be observed that the DFDMAC achieves the highest fairness index among all the MAC protocols. Introducing the proposed BT mechanism, IEEE 802.11ay can improve the fairness index by $32.58\%$, compared with the traditional one. This further proves that the proposed BT mechanism can effectively improve the fairness performance by overcoming deafness and directional HN problems. 

\subsection{Performance Evaluation of the Power Control Algorithm}
We consider a three-node FD link enabling simultaneous uplink and downlink transmissions between an AP and two user devices. The transmit power at a user device that enables uplink to AP varies from $1$ mW to $20$~mW with an interval of $1$~mW. The transmit power at AP that enables downlink to another user device varies from $1$~mW to $P_{AP}^{max}$ ($P_{AP}^{max} \in [20, 100]$~mW) with an interval of $1$~mW. The distance between the AP and each user device is $15$~m. The AP and each user device have $32$ and $8$ beams, respectively. Three MCS rates considered and the corresponding SINR thresholds are shown in Table \ref{MCS}. The SI cancellation level is set at $85$~dB. To account for potential IBI effects on SINR, we examine both the FD link with and without IBI. When power control is not utilized, transmitters always select maximum transmit power. Assume that uplink and downlink transmit DATA frames with identical payload size. Maximizing the FD link throughput turns into maximizing the received SINRs at the AP and the receiving user device. 

\begin{figure}[t]
    \graphicspath{{figure/}}
    \centering
    \subfigure[With IBI]{
        \includegraphics[width=4.1cm]{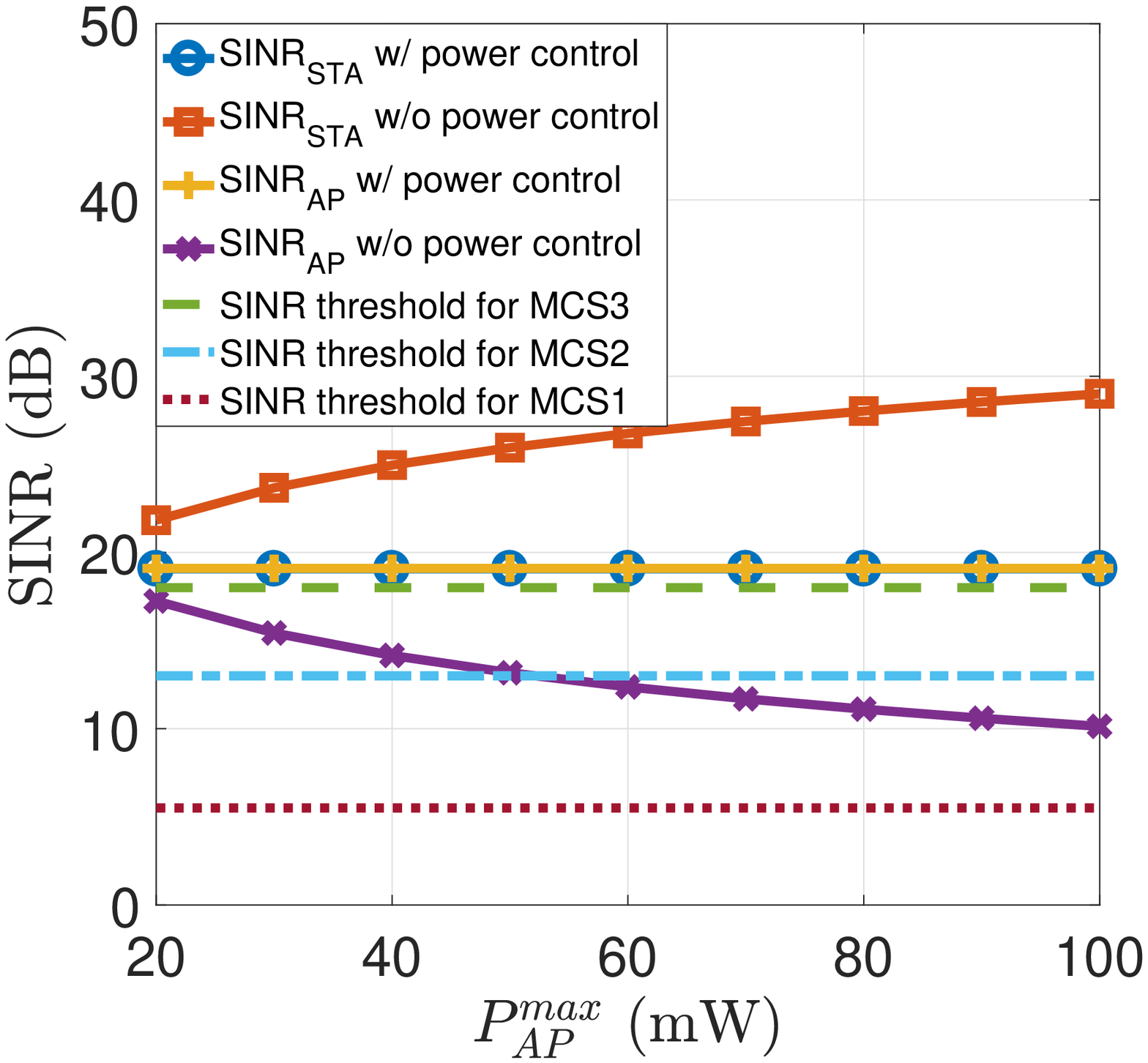}
    }
    \subfigure[Without IBI]{
        \includegraphics[width=4.1cm]{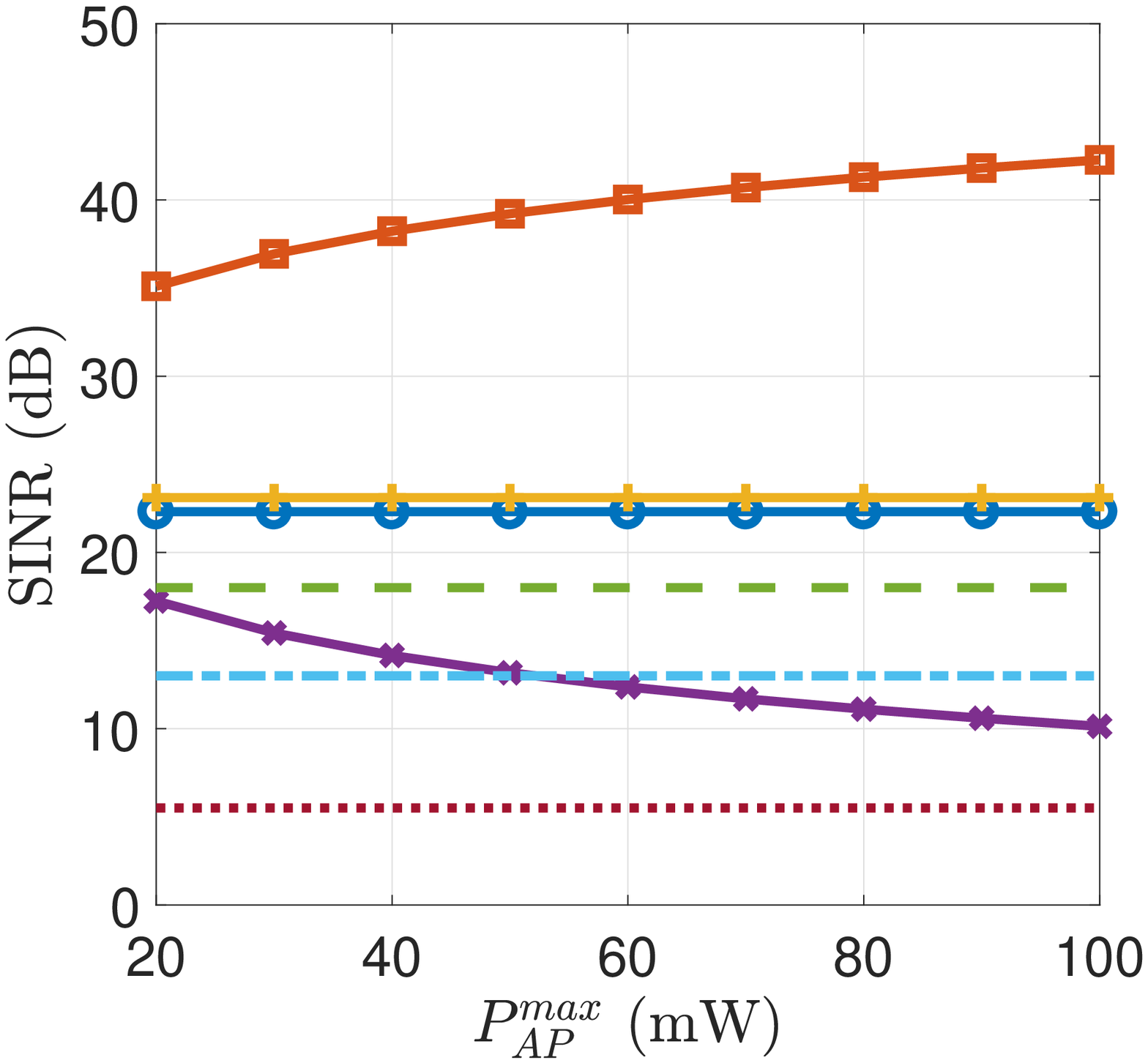}
    }
    \caption{The impact of power control on the SINRs of simultaneous uplink and downlink transmission.} \label{fig_SINR}
\end{figure}

Figure \ref{fig_SINR} shows the received SINRs at the AP and receiving user device with respect to $P_{AP}^{max}$. The two figures show that the SINR at the receiving user device increases, and the SINR at the AP decreases as $P_{AP}^{max}$ increases when the power control algorithm is not used. This is because the AP always selects the maximum transmit power to maximize the received SINR at the receiving user device, which leads to the decrease of the received SINR at the AP due to the impact of the residual SI. When power control is used to maximize the received SINRs of both uplink and downlink, it can be observed that both uplink and downlink can support MCS 3 in Fig. \ref{fig_SINR}(a) and Fig. \ref{fig_SINR}(b). Without power control, the transmission time of the FD link increases when the uplink only supports a lower transmission rate than the one using power control. Therefore, adjusting transmit power can effectively reduce the channel occupation time of an FD link and increase the link throughput.



\section{Conclusion}\label{conclusion}

In this paper, we have proposed a MAC protocol called DFDMAC for distributed mmWave FD networks. This protocol supports typical directional FD transmissions and can avoid deafness and directional HN problems with a designed novel busy-tone mechanism. To improve channel utilization and FD link throughput, we have designed a power control algorithm to obtain the optimal transmit power. Simulation results show that the DFDMAC can significantly improve the throughput and fairness performance. The proposed MAC protocol provides an effective solution for the deployment of distributed mmWave FD networks. In future work, we will extend the DFDMAC to overcome the blockage problem with FD amplify-and-forward transmission mode, which reduces communication latency.

\section*{Acknowledgment}

This work was supported in part by the Peng Cheng Laboratory Major Key Project under Grant PCL2021A09-B2 and by the Natural Science Foundation of China under Grant 6220012314.

\bibliographystyle{IEEEtran}
\bibliography{HNFD}

\end{document}